
\newcount\refnumber
\newcount\temp
\newcount\test
\newcount\tempone
\newcount\temptwo
\newcount\tempthr
\newcount\tempfor
\newcount\tempfiv
\newcount\testone
\newcount\testtwo
\newcount\testthr
\newcount\testfor
\newcount\testfiv
\newcount\itemnumber
\newcount\totalnumber
\refnumber=0
\itemnumber=0
\def\initreference#1{\totalnumber=#1
                 \advance \totalnumber by 1
                 \loop \advance \itemnumber by 1
                       \ifnum\itemnumber<\totalnumber
                        \temp=100 \advance\temp by \itemnumber
                        \count\temp=0 \repeat}

\def\ref#1{\temp=100 \advance\temp by #1
   \ifnum\count\temp=0
    \advance\refnumber by 1  \count\temp=\refnumber \fi
   \ [\the\count\temp]}

\def\reftwo#1#2{\tempone=100 \advance\tempone by #1
   \ifnum\count\tempone=0
   \advance\refnumber by 1  \count\tempone=\refnumber \fi
   \temptwo=100 \advance\temptwo by #2
   \ifnum\count\temptwo=0
   \advance\refnumber by 1  \count\temptwo=\refnumber \fi
 \testone=\count\tempone \testtwo=\count\temptwo
 \sorttwo\testone\testtwo
     \ [\the\testone,\the\testtwo]}       

\def\refthree#1#2#3{\tempone=100 \advance\tempone by #1
   \ifnum\count\tempone=0
    \advance\refnumber by 1  \count\tempone=\refnumber \fi
    \temptwo=100 \advance\temptwo by #2
   \ifnum\count\temptwo=0
    \advance\refnumber by 1  \count\temptwo=\refnumber \fi
    \tempthr=100 \advance\tempthr by #3
   \ifnum\count\tempthr=0
    \advance\refnumber by 1  \count\tempthr=\refnumber \fi
 \testone=\count\tempone \testtwo=\count\temptwo \testthr=\count\tempthr
 \sortthree\testone\testtwo\testthr
   \test=\testthr  \advance\test by -2
 \ifnum\test=\testone    \test=\testtwo  \advance\test by -1
    \ifnum\test=\testone   
    \ [\the\testone--\the\testthr]\fi \advance\temptwo by 1
  \else
     \ [\the\testone,\the\testtwo,\the\testthr]    
 \fi}

\def\reffour#1#2#3#4{\tempone=100 \advance\tempone by #1
   \ifnum\count\tempone=0
    \advance\refnumber by 1  \count\tempone=\refnumber \fi
    \temptwo=100 \advance\temptwo by #2
   \ifnum\count\temptwo=0
    \advance\refnumber by 1  \count\temptwo=\refnumber \fi
    \tempthr=100 \advance\tempthr by #3
   \ifnum\count\tempthr=0
    \advance\refnumber by 1  \count\tempthr=\refnumber \fi
    \tempfor=100 \advance\tempfor by #4
   \ifnum\count\tempfor=0
    \advance\refnumber by 1  \count\tempfor=\refnumber \fi
 \testone=\count\tempone \testtwo=\count\temptwo \testthr=\count\tempthr
 \testfor=\count\tempfor
 \sortfour\testone\testtwo\testthr\testfor
   \test=\testthr \advance\test by -1
   \ifnum\testtwo=\test   \test=\testtwo \advance\test by -1
    \ifnum\testone=\test  \test=\testfor \advance\test by -3
     \ifnum\testone=\test \ [\the\testone--\the\testfor]
     \else \ [\the\testone--\the\testthr,\the\testfor]
     \fi
    \else  \test=\testfor \advance\test by -1
     \ifnum\testthr=\test \ [\the\testone,\the\testtwo--\the\testfor]
     \else\ [\the\testone,\the\testtwo,\the\testthr,\the\testfor]
     \fi
    \fi
   \else \ [\the\testone,\the\testtwo,\the\testthr,\the\testfor]
   \fi}

\def\reffive#1#2#3#4#5{\tempone=100 \advance\tempone by #1
   \ifnum\count\tempone=0
    \advance\refnumber by 1  \count\tempone=\refnumber \fi
    \temptwo=100 \advance\temptwo by #2
   \ifnum\count\temptwo=0
    \advance\refnumber by 1  \count\temptwo=\refnumber \fi
    \tempthr=100 \advance\tempthr by #3
   \ifnum\count\tempthr=0
    \advance\refnumber by 1  \count\tempthr=\refnumber \fi
    \tempfor=100 \advance\tempfor by #4
   \ifnum\count\tempfor=0
    \advance\refnumber by 1  \count\tempfor=\refnumber \fi
    \tempfiv=100 \advance\tempfiv by #5
   \ifnum\count\tempfiv=0
    \advance\refnumber by 1  \count\tempfiv=\refnumber \fi
 \testone=\count\tempone \testtwo=\count\temptwo \testthr=\count\tempthr
 \testfor=\count\tempfor \testfiv=\count\tempfiv
 \sortfive\testone\testtwo\testthr\testfor\testfiv
  \test=\testthr \advance\test by -1
  \ifnum\testtwo=\test   \test=\testtwo \advance\test by -1
   \ifnum\testone=\test  \test=\testfor \advance\test by -3
    \ifnum\testone=\test \test=\testfiv \advance\test by -4
     \ifnum\testone=\test\ [\the\testone--\the\testfiv]
     \else\ [\the\testone--\the\testfor,\the\testfiv]
     \fi
    \else \ [\the\testone--\the\testthr,\the\testfor,\the\testfiv]
    \fi
   \else  \test=\testfor \advance\test by -1
    \ifnum\testthr=\test \test=\testfiv \advance\test by -2
     \ifnum\testthr=\test \ [\the\testone,\the\testtwo--\the\testfiv]
     \else \ [\the\testone,\the\testtwo--\the\testfor,\the\testfiv]
     \fi
    \else\ [\the\testone,\the\testtwo,\the\testthr,\the\testfor,\the\testfiv]
    \fi
   \fi
  \else \test=\testfor \advance\test by -1
   \ifnum\testthr=\test \test=\testfiv \advance\test by -2
   \ifnum\testthr=\test\ [\the\testone,\the\testtwo,\the\testthr--\the\testfiv]
    \else\ [\the\testone,\the\testtwo,\the\testthr,\the\testfor,\the\testfiv]
    \fi
   \else\ [\the\testone,\the\testtwo,\the\testthr,\the\testfor,\the\testfiv]
   \fi
  \fi}

\def\refitem#1#2{\temp=#1 \advance \temp by 100 \setbox\count\temp=\hbox{#2}}

\def\sortfive#1#2#3#4#5{\sortfour#1#2#3#4\relax
   \ifnum#5<#4\relax \test=#5\relax #5=#4\relax
     \ifnum\test<#3\relax #4=#3\relax
       \ifnum\test<#2\relax #3=#2\relax
         \ifnum\test<#1\relax  #2=#1\relax  #1=\test
         \else #2=\test \fi
       \else #3=\test \fi
     \else #4=\test \fi \fi}

\def\sortfour#1#2#3#4{\sortthree#1#2#3\relax
    \ifnum#4<#3\relax \test=#4\relax #4=#3\relax
       \ifnum\test<#2\relax #3=#2\relax
          \ifnum\test<#1\relax #2=#1\relax #1=\test
          \else #2=\test \fi
       \else #3=\test \fi \fi}

\def\sortthree#1#2#3{\sorttwo#1#2\relax
       \ifnum#3<#2\relax \test=#3\relax #3=#2\relax
          \ifnum\test<#1\relax #2=#1\relax #1=\test
          \else #2=\test \fi \fi}

\def\sorttwo#1#2{\ifnum#2<#1\relax \test=#2\relax #2=#1\relax #1=\test \fi}


\def\setref#1{\temp=100 \advance\temp by #1
   \ifnum\count\temp=0
    \advance\refnumber by 1  \count\temp=\refnumber \fi}

\def\printreference{\totalnumber=\refnumber
           \advance\totalnumber by 1
           \itemnumber=0
           \loop \advance\itemnumber by 1  
                 \ifnum\itemnumber<\totalnumber
                 \item{[\the\itemnumber]} \unhbox\itemnumber \repeat}

  \magnification=1200
  \hsize=15.5 truecm
  \vsize=23.0truecm
  \topskip=20pt            
  \fontdimen1\tenrm=0.0pt  
  \fontdimen2\tenrm=4.0pt  
  \fontdimen3\tenrm=7.0pt  
  \fontdimen4\tenrm=1.6pt  
  \fontdimen5\tenrm=4.3pt  
  \fontdimen6\tenrm=10.0pt 
  \fontdimen7\tenrm=2.0pt  
  \baselineskip=17.0pt plus 1.0pt minus 0.5pt  
  \lineskip=1pt plus 0pt minus 0pt             
  \lineskiplimit=1pt                           
  \parskip=2.5pt plus 5.0pt minus 0.5pt
  \parindent=15.0pt
\font\subsection=cmbx10 scaled\magstep1
\font\section=cmbx10 scaled\magstep2
\font\bx=cmr8
\def\lsim{\; \raise0.3ex\hbox{$<$\kern-0.75em\raise-1.1ex\hbox{$\sim$
}}\; }
\def\gsim{\; \raise0.3ex\hbox{$>$\kern-0.75em\raise-1.1ex\hbox{$\sim$
}}\; }
\def\jump{\vskip 1truecm}

\def\ns{\nu_{+}}

\def\nm{\nu_{\mu}}

If neutrinos are Dirac particles, then there will exist
right--handed chirality states of neutrinos $\nu_R$ and
left--handed chirality states of antineutrinos
$\overline\nu_L$ which in the Standard Model Lagrangian
do not have couplings to the weak interaction gauge bosons
W and Z.
The physical neutrino states are the helicity
eigenstates $\nu_{\pm}$ and $\overline\nu_{\pm}\ ,$ and  because
helicity and chirality eigenstates do not coincide
(only for ultra--relativistic neutrinos
${\nu}_{-}\simeq{\nu}_{L}$ and
${\nu}_{+}\simeq{\nu}_{R}$),
all the neutrino and antineutrino  helicity states
couple to the weak  gauge bosons.
Although the rate for W and Z mediated scattering between
ultrarelativistic fermions involving one $\nu_{+}$
($\overline\nu_{-}$) is suppressed, depending on the process at hand,
by a factor
$(10^{-2}-1)(m_\nu^2/2s)$, where $s$ is the centre--of--mass
energy squared, as compared with the corresponding reaction
involving $\nu_{-}$ ($\overline\nu_{+}$), the rate
might be sufficiently large to thermalize these 'wrong--helicity'
states in the early Universe.

If neutrinos with 'wrong' helicities should still be present
with an energy density comparable to $\nu_{-}$
at the time when neutron--to--proton ratio froze out
(at $T\simeq 0.7$ MeV), they would have
speeded up the expansion rate of the Universe (which depends on the
total energy density), with the consequence that the amount
of produced primordial $^4$He would have exceeded the current
observational limits. These imply that the effective number of
extra relativistic two--component neutrino species $\delta N_\nu$
is much less than one, with  $\delta N_{\nu}\lsim 0.3$ perhaps
the best limit\ref{1}. Thus the  production rate for $\nu_{+}$,
and hence their mass\ref{9},
must be small enough  so that they decouple  already before the
QCD phase transition which takes place at temperatures somewhere
between 100 and 400 MeV. In that case they will not participate in the entropy
transfer from quark--gluon plasma to particles in equilibrium,
and consequently their number and energy densities will be diluted
below levels that are acceptable for
primordial nucleosynthesis. The production rate was first estimated by
Fuller and Malaney\ref{2}, who argued that a
Dirac neutrino with a lifetime exceeding the nucleosynthesis
time scale $(t\sim{1}\ {\rm{s}})$ should have a mass less than about
300 keV.

This argument applies to neutrinos with a mass much less than 1 MeV.
A heavy neutrino with a mass in the MeV region would
have a more pronounced effect on nucleosynthesis than a light neutrino,
because  during the synthesis of the light elements the energy density
of the 'right--helicity' states of a heavy neutrino
would be comparable to or higher than that of a massless neutrino.
This is because at that epoch these states have already decoupled
($T^{\nu_{-}}_{dec}\sim$ few MeV). This has been shown\ref{3} to lead to
an excluded region
$0.5\ {\rm MeV}\lsim{m}_{\nu_\tau}\lsim 30\ {\rm MeV}$ for the tau neutrino
mass, provided $\tau_{\nu_\tau}\gsim{10^3}$ s.
(If $1\ {\rm s}\lsim\tau_{\nu_\tau}\lsim 10^3$ s,
the upper bound is somewhat weakened.)

These considerations are  relevant not only for the tau
neutrino mass, the laboratory limit on which is
$m_{\nu_\tau}< 31$ MeV\ref{4}, but possibly also for
the muon neutrino. This is because the exprimental $\nm$ mass
limit has recently been revised upward\ref{5},
with the current limit being about $m_{\nm} <500$ keV.

The original Fuller and Malaney limit was based on an
approximate estimate for the rate of $\nu_+$ production.
In this Letter we shall present a careful re--evaluation of this
limit by computing all the relevant cross sections and decay rates exactly.
As we shall show by explicit calculation,
the mass limit on $\nu_\tau$ ($\nu_\mu$) becomes larger
as much as by a factor of four (two) as compared with the
Fuller and Malaney result.

Before the QCD phase transition but below, say,
$T\simeq 0.5$ GeV, the fermions present in the Universe at
significant number densities were the leptons
and u, d, s and c quarks.
All the $2\rightarrow 2$ scattering processes involving them,
with no 'wrong--helicity' neutrinos in the initial state
and with at least one
$\nu_+^{\mu}$ or $\nu_+^{\tau}$
in the final state,  are listed in Table 1. (There are altogether 47
separate reactions for each of the neutrinos which need to be taken
into consideration).
The first constraint is imposed on because each wrong helicity
neutrino in the initial (final) state introduces an additional
small factor $m_\nu^2/|{\bf p}|^2$
($m_\nu^2/|{\bf p}^{\prime}|^2$) to the cross section.
Here $|{\bf p}|$ and $|{\bf p}^{\prime}|$ are the absolute
values of the centre--of--mass momenta of the incoming
and outgoing particles, respectively.
Hence processes with more than one 'wrong--helicity'
neutrino can be ignored as compared to prosesses with only
one 'wrong--helicity' neutrino.

We have computed the cross sections for these scatterings
in the limit of low momentum transfer
$|q^2|\ll M_Z^2,\ M_W^2,\ M_H^2$.
The contributions to the cross sections arising from the exchange of
gauge bosons are given by
$$
\sigma_{+}^{(12\rightarrow{34})}=
{G_F^2\over\pi}{|{\bf p}||{\bf p}^{\prime}|^3\over s}
F_{+}^{(12\rightarrow{34})}
\left ({m_1^2\over |{\bf p}|^2},{m_2^2\over |{\bf p}|^2},
{m_3^2\over |{\bf p}^{\prime}|^2},
{m_4^2\over |{\bf p}^{\prime}|^2}\right ),
\eqno(1)
$$
where we have summed
over all helicity and colour states of all charged leptons and
quarks. The dimensionless functions $F_{+}^{(12\rightarrow{34})}$
are listed in Table 1. In Eq.\ (1) the coefficient of
$F_{+}^{(12\rightarrow{34})}$ is of the order
of an ordinary weak interaction cross section for
fermion-fermion scattering,
so that all suppression due to the 'wrong--helicity'
neutrino in the final state is included
in this function. For a massless neutrino
$F_{+}^{(12\rightarrow{34})}=0$,
and $F_{+}^{(12\rightarrow{34})}$ increases
monotonically as a function of its arguments.
Some of the processes, such as
$\nu^k_{-}\nu^k_{-}\rightarrow{\nu^k_{+}\nu^k_{+}}$ and
$\nu^k_{-}\nu^j_{-}\rightarrow{\nu^k_{+}\nu^j_{+}}$
$(k=\mu,\tau\neq j=e,\mu,\tau )$
require two mass insertions and are therefore suppressed. Similar
suppression holds in the relativistic limit also
for the charged--current processes
$l^{k-}\nu^j_{-}\rightarrow\nu^k_{+}l^{j-}$ and
$l^{k-}u^m\rightarrow\nu^k_{+}d^n$ ($u^m=u,c;\ d^n=d,s$),
despite the fact that there is
only one 'wrong--helicity' neutrino.

Having exact formulas one can estimate the cross sections
for 'helicity--flip' scattering in the relativistic limit.
It turns out that the popular approximation
$\overline{\sigma}_{+}\approx
{G_F^2}{E_\nu^2}(m_{\nu}/2E_\nu)^2$,
where the overbar indicates the averaging over the helicity
and colour states of the initial charged leptons and quarks,
overestimates the
cross section for all the processes typically by more than an order
of magnitude.

Neutral current scattering between fermions can
also be mediated by the Higgs boson H.
Although the Yukawa vertices provide direct 'helicity--flip'
interactions for fermions, the fermion couplings to Higgs
are weaker than gauge couplings roughly by a factor $m_f/M_W$.
In the low energy
limit additional suppression would arise from the
propagators if the mass of the Higgs is larger than
$M_W$ and $M_Z$. We have calculated the contributions from the Higgs
boson exchange and from the interference
between gauge boson(s) and the Higgs boson, and found that
these are smaller at least by a factor  $|{\bf{p}}^{\prime}|^2/M_H^2$
than gauge boson contributions to the cross section.
In what follows we shall neglect the Higgs boson contribution.

Besides the scattering processes, $\nu_\mu$ and $\nu_\tau$
can be created also in three-body decays. There exist six relevant
decay channels producing $\nu_+^\mu$:
$$\eqalign{
\mu^{-}&\rightarrow\nu_+^{\mu}{e^{-}}\,\overline{\nu}_e,\cr
\mu^{-}&\rightarrow\nu_+^{\mu}{d}\,\overline{u},\cr
c&\rightarrow{s}\,\mu^+\nu_+^{\mu},\cr
c&\rightarrow{d}\,\mu^+\nu_+^{\mu},\cr
\overline{s}&\rightarrow\overline{u}\,\mu^+\nu_+^{\mu},\cr
\tau^{+}&\rightarrow\overline{\nu}^{\tau}\mu^+\nu_+^{\mu},\cr
}\eqno(2)
$$
and another six decay channels producing $\nu_+^\tau$:
$$\eqalign{
\tau^{-}&\rightarrow\nu_+^{\tau}{e^{-}}\overline{\nu}_{e},\cr
\tau^{-}&\rightarrow\nu_+^{\tau}{\mu^{-}}\overline{\nu}_{\mu},\cr
\tau^{-}&\rightarrow\nu_+^{\tau}{d}\,\overline{u},\cr
\tau^{-}&\rightarrow\nu_+^{\tau}{d}\,\overline{c},\cr
\tau^{-}&\rightarrow\nu_+^{\tau}{s}\,\overline{c},\cr
\tau^{-}&\rightarrow\nu_+^{\tau}{s}\,\overline{u}.\cr
}\eqno(3)
$$
The decay rate for particles,
e.g. $\mu^{-}\rightarrow\nu_+^{\mu}{e^{-}}\,\overline{\nu}_e$, is in the
rest frame  given by
$$\eqalign{&\cr
\Gamma_+^{(1\rightarrow{234})}=&
{{G_F^2}\over{12\pi^3}}\int_0^{|{\bf p}_4|_{max}}
d|{\bf p}_{4}|
{{|{\bf p}_{4}|^{2}(\sqrt{|{\bf p}_{4}|^{2}+m_4^2}-|{\bf p}_{4}|)}
\over{\sqrt{|{\bf p}_{4}|^{2}+m_4^2}}}
{{{\Delta}^{1/2}(m_{14}^2,m_2^2,m_3^2)}\over{m_{14}^4}}\cr
&\Bigl\{{\Delta}(m_{14}^2,m_2^2,m_3^2)
+2{{[m_{14}^4+(m_2^2+m_3^2)m_{14}^2-2(m_2^2-m_3^2)^2]}
\over{m_{14}^2}}\cr
&\times{{(m_1-\sqrt{|{\bf p}_{4}|^{2}+m_4^2})
(m_1\sqrt{|{\bf p}_{4}|^{2}+m_4^2}-m_1|{\bf p}_{4}|-m_4^2)}
\over{(\sqrt{|{\bf p}_{4}|^{2}+m_4^2}-|{\bf p}_{4}|)}}\Bigr\},\cr &\cr}
\eqno(4a)
$$
while the decay rate for antiparticles,
e.g. $\tau^{+}\rightarrow\overline{\nu}_{\tau}\mu^+\nu_+^{\mu}$,
is given by
$$\eqalign{&\cr
\Gamma_+^{(1\rightarrow{234})}=
&{{G_F^2}\over{6\pi^3}}\int_0^{|{\bf p}_4|_{max}}
d|{\bf p}_{4}|
{{|{\bf p}_{4}|^{2}(\sqrt{|{\bf p}_{4}|^{2}+m_4^2}-|{\bf p}_{4}|)}
\over{\sqrt{|{\bf p}_{4}|^{2}+m_4^2}}}
{{{\Delta}^{1/2}(m_{14}^2,m_2^2,m_3^2)}\over{m_{14}^4}}\cr
&\bigl\{2{\Delta}(m_{14}^2,m_2^2,m_3^2)
+[m_{14}^4+(m_2^2+m_3^2)m_{14}^2-2(m_2^2-m_3^2)^2]\bigr\},\cr & \cr}
\eqno(4b)
$$
with
$m_{14}^2\equiv{m}_1^2-2m_{1}\sqrt{|{\bf p}_{4}|^{2}+m_4^2}+m_4^2$
and ${\Delta}(a,b,c)\equiv{a}^2+b^2+c^2-2ab-2bc-2ca$.
Here the subscript $4$ refers to $\nu_{+}^\mu$ or $\nu_{+}^\tau$ and
the maximal momentum of the 'wrong--helicity' neutrino is
$|{\bf p}_4|_{max}=
\sqrt{[m_1^2-(m_2+m_{3})^2+m_4^2]^2-4{m_1}^2{m_4}^2}/2{m_1}$.
The expressions for particle and antiparticle decay rates
are different because they do not correspond to CP--conjugated
processes: in both cases the final state involves the
'wrong--helicity' neutrino (but not anti--neutrino).
In these rates we have summed over all helicity states of all
particles except the $\nu_{+}^{\mu{(}\tau{)}}$ under consideration.
If quarks are included in the decay process these
decay rates must be multiplied by a factor $3V_{mn}^2$ which
accounts for the three colour states of quarks and for quark mixing.

The thermally averaged scattering rate reads \ref{6}
$$
\Gamma_{+}^{sc}=
{1\over{n_{\nu_+}^{eq}(T)}}\sum_{(12\rightarrow{34})}
\int{d^3{p}_1\over{(2\pi)^3}}{d^3{p}_2\over{(2\pi)^3}}
f(E_1/T)f(E_2/T){\sigma}_{+}^{(12\rightarrow{34})}j(p_1,p_2),
\eqno(5)
$$
where $n_{\nu_+}^{eq}$ is the equilibrium number density of
$\nu_+$'s, $f(E_i/T)$
are the Fermi--Dirac distributions of the incoming
particles, and
$j(p_1,p_2)=\sqrt{(p_1\cdot p_2)^2-m_1^2m_2^2}/E_1E_2$ is a
flux--related factor.
In (5) we have neglected the final state Pauli blocking,
which is an about 10\%\ effect \ref{6}.

Thermally averaged decay rate is simply given by
$$
\Gamma_{+}^d=
{1\over{n_{\nu_+}^{eq}(T)}}\sum_{(1\rightarrow{234})}
\Gamma_{+}^{(1\rightarrow{234})}
\int{d^3{p}_1\over{(2\pi)^3}}
f(E_1/T){{m_1}\over{E_1}},
\eqno(6)
$$
where the factor $m_1/E_1$ arises from the Lorentz boost
of the decay rate.

We have estimated the total thermally averaged $\nu_{+}$ production rate
$\Gamma_+=\Gamma_+^{sc}+\Gamma_+^d$ numerically, and the result is
displayed in Fig.\ 1.
The difference between the production rates for $\nu_{+}^\mu$ and
$\nu_{+}^\tau$ is due to the differences in the phase spaces (initial and/or
final) of the corresponding processes. While $\mu$ is relativistic in the
range of temperatures $100-400\ {\rm MeV}$, $\tau$ is
nonrelativistic and therefore its number density, being
Boltzmann--suppressed, is very sensitive to changes in temperature. This
can be illustrated by the fact that while
at $T\approx 100$ MeV purely charged processes do not contribute
to the $\nu_{+}^\tau$ production, their contribution
is about 40\%\ at $T\approx 250$ MeV and about $75\%$ at $T\approx 400$ MeV.
In contrast, for $\nu_{+}^\mu$ this figure rises from $55\%$
($T\approx 100$ MeV) to $70\%$ ($T\approx 400$ MeV). One can see that
elastic scattering and
annihilation processes, dominated by
$\nu^k_{-}\overline\nu^k_{+}\rightarrow\nu^k_{+}\overline\nu^k_{+}$,
are not so important as one would naively expect.
Comparing with the Fuller and Malaney result, which is also shown in
Fig.\ 1, we find that at $T\approx{100}$ MeV the actual rate
for $\nu^\tau_{+}\ (\nu^\mu_{+})$ production is $18\ (7)$ times smaller.

We now require that the production of $\ns$'s ceases latest at
the onset of the QCD phase transition, or that
$$
\Gamma_{+}(m_{\nu},T_{\rm{QCD}})\le H(T_{\rm{QCD}}),
\eqno(7)
$$
where the Hubble parameter $H$ is given by
$$
H=\sqrt{{8\pi\over 3M_{Pl}^2}\rho(m_i,T)}\equiv
\sqrt{{4\pi^3g_{e\!f\!f}(m_i,T)\over 45}}{T^2\over M_{Pl}}.
\eqno(8)
$$
Here $\rho(m_i,T)$ is the total energy density, including also
the particles that could be non--relativistic at the time of QCD
phase transition, such as $\tau$, $c$, $s$ and $\mu$.
Here we differ from the treatment of Fuller and Malaney, who
did not account for the non--relativistic degrees of freedom in
their estimate. Thus
$g_{e\!f\!f}$ counts all the degrees of freedom, and
we have tabulated it in Table 2, assuming that the
non--relativistic species stay in equilibrium with the appropriate
Boltzmann suppressed equilibrium densities.
We have used the current quark masses
$m_u\approx 5.6\pm 1.1$ MeV, $m_d\approx 9.9\pm 1.1$ MeV,
$m_s\approx 199\pm 33$ MeV, and $m_c\approx 1.35\pm 0.05$ GeV \ref{7}.
About the half of the difference between the minimal and maximal
values of $g_{e\!f\!f}$ is due to the uncertainty in quark masses.
Another half comes from the uncertainty as to whether both or only one
of the neutrinos $\nu_{+}^\mu$ and $\nu_{+}^\tau$ is in equilibrium.

As the production rate for the 'wrong--helicity' neutrino
increases together with its mass (at $T\gg{m_\nu}\ \
\Gamma_{+}\propto{m_\nu^2}$), and for the temperatures and masses under
consideration H does not depend on $m_\nu$, there exists
an upper limit on the neutrino mass at which the inequality (7) can be
satisfied. These mass limits for $\nu_\tau$ and
$\nu_\mu$ as functions of the QCD phase transition temperature
are displayed in Fig. 2. The value of the upper limit
of the neutrino mass is almost
insensitive to the variations of the quark masses within ranges
given above, despite the fact that the number and energy densities of
s and c quarks change considerably. This happens because
the changes in the production rate and the expansion rate
compensate for each other.
{}From Fig. 2 one can readily see that if
$T_{\rm{QCD}}\simeq{(100)200}$ MeV,
the mass limits are $m_{\nu_\tau}\lsim (1180)740$ keV and
$m_{\nu_\mu}\lsim{(720)480}$ keV. According to ref.\ref{3}, however,
such large masses are already in conflict with primordial
nucleosynthesis. Thus we may conclude that the equilibration
of 'wrong' Dirac neutrino helicity states does not yield any
additional new limit.
Note that because the dilution provided by the entropy production at the QCD
phase transition is more than sufficient for nucleosynthesis, $\ns$'s
could well decouple some time during the actual phase transition, lowering
the decoupling temperature and increasing the cosmological neutrino mass
limit. Assuming decoupling at the onset of the phase transition will
therefore slightly underestimate the actual upper limit.
\vfill\eject\null\jump
\centerline{\subsection References}\jump
\refitem{1}{T.P. Walker {\sl et al.}, Astrophys.\ J.\ {\bf 376} (1991) 393.
}
\refitem{2}{G.M.\ Fuller and R.A.\ Malaney, Phys.\ Rev.\ {\bf D43}
(1991) 3136.}
\refitem{3}{E.W. Kolb {\sl et al.}, Phys.\ Rev. Lett.\ {\bf 67} (1991) 533.
}
\refitem{4}{H.\ Albrecht {\sl et al.}\ (ARGUS collaboration), Phys.\ Lett.\
{\bf B292} (1992) 221.}
\refitem{5}{R.G.H.\ Robertson, Talk given in XXVI Int. Conf. on High
Energy
Physics.}
\refitem{6}{K.\ Enqvist, K.\ Kainulainen and V.\ Semikoz,  Nucl.\
 Phys.\ {\bf B374} (1992) 392.}
\refitem{7}{Particle Data Group, K.\ Hikasa {\sl et al.},
{\sl Review of Particle Properties}, Phys.\ Rev.\ {\bf D45} (1992).}
\refitem{9}{K.J.F.\ Gaemers, R.\ Gandhi and J.M.\ Lattimer, Phys.\ Rev.\
{\bf D40} (1989) 309.}
\printreference
\vfill\eject

\def\jump{\vskip 1truecm}

\hoffset=0truecm

{\settabs 4\columns
\+\hrulefill & \hrulefill & \hrulefill & \hrulefill & \cr
{\vskip -19pt}
\+\hrulefill & \hrulefill & \hrulefill & \hrulefill & \cr
{\vskip 7pt}
\+{Process}\hfill &
$F_{+}^{(12\rightarrow{34})}(x_1,x_2,y_3,y_4)$ & &\cr
\+\hrulefill & \hrulefill & \hrulefill & \hrulefill & \cr
\+ $\nu^k_{-}\nu^k_{-}\rightarrow\nu^k_{+}\nu^k_{+}$ & $x_{\nu^k}^2$ & &\cr
\+ $\nu^k_{-}\overline\nu^k_{+}\rightarrow\nu^k_{+}\overline\nu^k_{+}$
&${(1/3)}x_{\nu^k}(\sqrt{1+x_{\nu^k}}+1)^2$ & &\cr
\+ $\nu^k_{-}\nu^j_{-}\rightarrow\nu^k_{+}\nu^j_{+}$
&${(1/2)}x_{\nu^k}x_{\nu^j}$ & &\cr
\+ $\nu^k_{-}\overline\nu^j_{+}\rightarrow\nu^k_{+}\overline\nu^j_{+}$
& ${(1/12)}x_{\nu^k}(\sqrt{1+x_{\nu^j}}+1)^2$ & &\cr
\+ $\nu^k_{-}l^{k\mp}\rightarrow\nu^k_{+}l^{k\mp}$
&${(1/6)}x_{\nu^{k}}
[2(c^{\prime{l}}_V\mp{c^{\prime{l}}_A})^{2}
+3(c^{\prime{l2}}_V+3c^{\prime{l2}}_A)x_{l^k}]$ & &\cr
\+ $\nu^k_{-}l^{j\mp}\rightarrow\nu^k_{+}l^{j\mp}$
&${(1/6)}x_{\nu^k}[2(c^l_V\mp{c^l_A})^{2}+3(c^{l2}_V+3c^{l2}_A)x_{l^j}]
$ & &\cr
\+ $
\nu^k_{-}q\rightarrow\nu^k_{+}q
$
&$
{(1/2)}x_{\nu^k}
[2(c^q_V-{c^q_A})^{2}+3(c^{q2}_V+3c^{q2}_A)x_q]
$& &\cr
\+ $
\nu^k_{-}\overline{q}\rightarrow\nu^k_{+}\overline{q}
$&$
{(1/2)}x_{\nu^k}
[2(c^q_V+{c^q_A})^{2}+3(c^{q2}_V+3c^{q2}_A)x_q]
$& &\cr

\+ $
\nu^j_{-}\overline{\nu}^j_{+}\rightarrow\nu^k_{+}\overline{\nu}^k_{+}
$&$
{(1/12)}y_{\nu^k}(\sqrt{1+x_{\nu^j}}+1)^2
$& &\cr

\+ $
l^{k-}l^{k+}\rightarrow\nu^k_{+}\overline{\nu}^k_{+}
$&$
{(1/3)}y_{\nu^k}(c^{\prime{l2}}_V+c^{\prime{l2}}_A)(2+3x_{l^k})
$& \cr

\+ $
l^{j-}l^{j+}\rightarrow\nu^k_{+}{\overline{\nu}}^k_{+}
$&$
{(1/3)}y_{\nu^k}(c^{l2}_V+c^{l2}_A)(2+3x_{l^j})
$& &\cr

\+ $
q\overline q\rightarrow\nu^k_{+}\overline\nu^k_{+}
$&$
y_{\nu^k}(c^{q2}_V+c^{q2}_A)(2+3x_q)
$ & &\cr

\+ $
l^{k-}\nu^j_{-}\rightarrow\nu^k_{+}l^{j-}
$&$
2(\sqrt{1+y_{\nu^k}}-1)(\sqrt{1+x_{l^k}}+1)
(\sqrt{1+x_{\nu^j}}+1)(\sqrt{1+y_{l^j}}-1)
$& &\cr

\+ $
\nu^j_{-}l^{j+}\rightarrow\nu^k_{+}l^{k+}
$&$
{(2/3)}(\sqrt{1+y_{\nu^k}}-1)(\sqrt{1+x_{\nu^j}}+1)
(3\sqrt{1+x_{l^j}}\sqrt{1+y_{l^k}}-1)
$& &\cr

\+ $
l^{k-}l^{j+}\rightarrow\nu^k_{+}\overline{\nu}^j_{+}
$&$
{(2/3)}(\sqrt{1+y_{\nu^k}}-1)(\sqrt{1+y_{\nu^j}}+1)
(3\sqrt{1+x_{l^k}}\sqrt{1+x_{l^j}}-1)
$& &\cr

\+ $
l^{k-}u^m\rightarrow\nu^k_{+}d^n
$&$
12{V}_{mn}^{2}(\sqrt{1+y_{\nu^k}}-1)(\sqrt{1+y_{d^n}}-1)
(\sqrt{1+x_{l^k}}\sqrt{1+x_{u^m}}+1)
$& &\cr

\+ $
l^{k-}\overline{d}^n\rightarrow\nu^k_{+}\overline{u}^m
$&$
4{V}_{mn}^{2}(\sqrt{1+y_{\nu^k}}-1)
(3\sqrt{1+x_{l^k}}\sqrt{1+x_{d^n}}\sqrt{1+y_{u^m}}-1)
$& &\cr

\+ $
u^m\overline{d}^n\rightarrow\nu^k_{+}l^{k+}
$&$
4{V}_{mn}^{2}(\sqrt{1+y_{\nu^k}}-1)
(3\sqrt{1+x_{u^m}}\sqrt{1+x_{d^n}}\sqrt{1+y_{l^k}}-1)
$ & &\cr
\+\hrulefill & \hrulefill & \hrulefill & \hrulefill & \cr}

\jump\noindent
{\bf Table 1.}
List of $2\rightarrow 2$ scattering cross sections for $\nu_+^k$ production.
$F_{+}^{(12\rightarrow{34})}$ is defined by Eq.\ (1), and the notations are:
$k=\mu ,\; \tau\neq j=e,\; \mu ,\; \tau ;\ u^m=u,\; c;\ d^m=d,\; s$;
$x_f\equiv{{m_f^2}/{|{\bf p}|^{2}}}$ and
$y_f\equiv{{m_f^2}/{|{\bf p}^{\prime}|^{2}}}$;
$c^{\prime{f}}_V\equiv{c^f_V}+1$,
$c^{\prime{f}}_A\equiv{c^f_A}+1$,
$c^f_V\equiv{T}^{3f}_L-2Q^{f}{\sin}^{2}{\theta}_W$ and
$c^f_A\equiv{T}^{3f}_L$
are the vector and axial vector couplings;
$V_{mn}$ is an element of the C-K-M matrix.
We have used the values $\sin^{2}{\theta}_W=0.2325$,
$V_{ud}=0.9753$, $V_{us}=0.221$, $V_{cd}=0.221$ and
$V_{cs}=0.9743$ \ref{7}. The present experimental uncertanties
in these values do not affect our upper limits on the
neutrino masses.
\vfill\eject
$$
\vbox{\tabskip=0pt
\halign to 175pt
{{\tabskip=3em plus 5em minus 5em}
#& \hfil#\hfil & \hfil#\hfil & \hfil#\hfil &#{\tabskip=0pt} \cr
\noalign{\hrule} \cr
\noalign{\vskip -2pt \hrule \vskip 7pt} \cr
& $T$(MeV) & $g_{e\!f\!f}^{\rm min}$ & $g_{e\!f\!f}^{\rm max}$ & \cr
\cr\noalign{\hrule} \cr
& 100 & 59.4 & 62.8 & \cr
& 150 & 61.6 & 64.3 & \cr
& 200 & 62.8 & 65.3 & \cr
& 250 & 64.0 & 66.5 & \cr
& 300 & 65.3 & 67.8 & \cr
& 350 & 66.7 & 69.1 & \cr
& 400 & 67.9 & 70.3 & \cr
\cr\noalign{\hrule} \cr
}}$$
{\vskip 0.4truecm}\noindent
{\bf Table 2.}
The effective number of degrees of freedom as a
function of temperature.
\vfill\eject
\hoffset=0truecm\null\jump
\centerline{\subsection Figure Captions}\jump
\noindent{\bf Fig.~1.} {The total thermally averaged production rate
for $\nu^\mu_+$ and $\nu^\tau_+$. For comparison, we show also
the Fuller and Malaney rate (dashed curve).}
\jump
\noindent{\bf Fig.~2.} {Neutrino mass limits corresponding to
$\Gamma_+\le H$. The allowed region is below the curves.}
\vfill\eject
\null\nopagenumbers\vskip -20pt
\line {\hfill NORDITA--92/87 P}
\vskip 1.2truecm
\centerline {\subsection Cosmological neutrino mass limit revisited}
\vskip 1.5truecm
\centerline {Kari Enqvist{\footnote{\bx *}{\bx e-mail:
enqvist@nbivax.nbi.dk}}
and Hannes Uibo{\footnote{\bx $^\dagger$}{\bx e-mail:
uibo@nbivax.nbi.dk; on leave of
absence from Inst. of Physics, Estonian Academy of \vskip-7pt
Sciences, Riia 142, EE2400 Tartu, Estonia.}}}
\vskip 0.4truecm
\centerline {Nordita, Blegdamsvej 17, DK-2100 Copenhagen \O,
Denmark}
\vskip 2.0truecm
\centerline {\bf Abstract}
\vskip 1truecm\noindent
We consider the equilibration of the 'wrong--helicity' Dirac neutrino
states $\ns$ and $\overline\nu_-$ in the early Universe via weak
interactions and calculate
carefully the thermally averaged production rate, taking into account all
the relevant scattering and decay processes.
Requiring that the production rate is less than the Hubble
parameter at the onset of QCD phase transition so that the
nucleosynthesis predictions are not contradicted, we find  for
$T_{\rm{QCD}}\simeq 200$ MeV
the upper limits $m_{\nu_\tau}\lsim 740$ keV and
$m_{\nu_\mu}\lsim 480$ keV.

\end